\documentclass[11pt]{article}

\begin{document}

\title{V.M. Red'kov\footnote{redkov@dragon.bas-net.by} \\
Particle with spin  $S=3/2$  in Riemannian space-time\footnote{ Chapter 7   in: V.M. Red'kov,
Fields in Riemannian space  and the Lorentz group (in Russian).
Publishing House "Belarusian Science", Minsk, 2009.}\\
{\small  Institute of Physics, National Academy of Sciences of
Belarus}
}

\date{}

\maketitle
\begin{abstract}

Equations for   16-component vector-bispinor field, originated from Rarita--Schwinger Lag\-ran\-gian for spin 3/2 field
 extended
to Riemannian space-time are investigated.
Additional general covariant constrains for the field are produced, which for
some space-time models greatly simplify original wave equation.

Peculiarities in description of the massless spin 3/2 field are specified.
In the flat Minkowski space for massless case there exist gauge invariance of the main wave equation,
which reduces to possibility to produce a whole  class of trivial solutions in the the form  of 4-gradient
of arbitrary(gauge)  bispinor function,  $\Psi ^{0}_{c}(x) =
\partial_{c}  \psi (x) $.
Generalization of that property for  Riemannian model is performed; it is shown that
in general covariant case solutions of the gradient type
$\Psi ^{0}_{\beta }(x) =  ( \nabla _{\beta }\; + \; \Gamma _{\beta })
 \Psi (x) $ exist in space-time regions where the Ricci tensor  obeys  an  identity
 $ R_{\alpha \beta } -  {1 \over 2}R  g_{\alpha \beta }  = 0
 $.

\end{abstract}

\section{Massive field, additional constraints }

Starting with  fundamental investigation by  Pauli and Fierz
\cite{1939-Pauli(2), 1939-Fierz-Pauli}, also Rarita and Schwinger  
\cite{1941-Rarita},  the field with spin $3/2$ always attracted attention:

\begin{quotation}

 Ginzburg  \cite{1942-Ginzburg},
 Davydov \cite{1943-Davydov}, Tamm -- Davydov -- Ginzburg
  \cite{1947-Tamm(1), 1947-Tamm(2)}, Ginzburg -- Smorodinskiy \cite{1943-Ginzburg(1),  1943-Ginzburg(2)},
 Fradkin  \cite{1950-Fradkin},
Belinfante  \cite{1953-Belinfante},
 Fainberg \cite{1955-Fainberg},
 Petras  \cite{1955-Petras(1), 1955-Petras(2)},
Jonson and Sudarshan   \cite{1961-Johnson(1), 1961-Johnson(2)},  Bender and McKoy \cite{1966-Bender},
Munczek  \cite{1967-Munczek}, Velo and Zwanziger \cite{1969-Velo(1), 1969-Velo(2)},
Hagen and Singh  \cite{1971-Hagen,   1973-Singh, 1974-Hagen, 1974-Singh-Hagen, 1982-Hagen},
 Baisya \cite{1971-Baisya},  Fisk and Tait  \cite{1973-Fisk},
  Hortacsu   \cite{1974-Hortacsu},
Mathews et al  \cite{1975-Mathews(1), 1975-Mathews(2)},
Madore  and Tait \cite{1973-Madore, 1975-Madore},
Hasumi, Endo and Kimura \cite{1975-Hasumi-Endo-Kimura},
 Lopes -- Spehler -- Leite -- Fleury      \cite{1979-Lopes, 1982-Leite},
Aurilia et al \cite{1980-Aurilia},
Inoue -- Omote -- Kobayash \cite{1980-Inoue-Omote-Kobayash},
 Loide  \cite{1984-Loide, 1986-Loide},
 Pletjuxov and Strazhev  \cite{1985-Pletjuxov},
  Labonte  \cite{1980-Labonte, 1981-Labonte},
Capri and  Kobes     \cite{1980-Capri}, Barut and Xu  \cite{1982-Barut},
Darkhosh   \cite{1985-Darkhosh},  Rindani and  Sivakumar
\cite{1986-Rindani},
 Cox \cite{1989-Cox},
  Penrose \cite{1991-Penrose},
  Pascalutsa \cite{1998-Pascalutsa},
 Haberzettl  \cite{1998-Haberzettl},
Deser S.,  Waldron A.,  Pascalutsa
  \cite{2000-Deser, 2001-Deser(2)},
Kirchbach and  Ahluwalia \cite{2001-Ahluwalia},
 Gsponer and Hurni  \cite{2003-Gsponer}, Pilling
\cite{2004-Pilling(1), 2004-Pilling(2)},  Kaloshin and Lomov \cite{2004-Lomov, 2011-Lomov},
Napsuciale -- Kirchbach -- Rodriguez
\cite{Napsuciale-Kirchbach-Rodriguez}.

\end{quotation}

On the curved space-time background, the field of spin $3/2$ is investigated  much less
than fields of spins  $0, 1/2, 1$. This circumstance is due to
complexity of this object: all its 16 components are tightly linked to each other by
presence of curved geometry.
Let us consider some peculiarities in description of the particle with spin  $3/2$ in Riemannian space-time,
first specifying the massive case.

Lagrangian by Rarita -- Schwinger extended to generally covariant case has the form
(let it be  $k = imc / \hbar $)

$$
L =   {1 \over 2}  \; [ \;\bar{\Psi }_{\alpha } \gamma ^{\beta
}(x) \stackrel{\rightarrow}{D}_{\beta }  \Psi ^{\alpha } -
\bar{\Psi }_{\alpha } \gamma ^{\beta }(x)
\stackrel{\leftarrow}{D}_{\beta }  \Psi ^{\alpha } \; ]
$$

$$
+ {1 \over 3}  \; [\;  \bar{\Psi }_{\alpha } \gamma ^{\alpha } (x)
\stackrel{\rightarrow}{D}_{\beta } \Psi ^{\beta } \; - \;
\bar{\Psi }_{\alpha } \gamma ^{\alpha }(x)
\stackrel{\leftarrow}{D} _{\beta } \Psi  ^{\beta } \; ]
$$

$$
+ {1 \over 6}  \; [\;  \bar{\Psi }_{\alpha } \gamma ^{\alpha }(x)
\gamma ^{\beta } \stackrel{\rightarrow}{D}_{\beta } \gamma ^{\rho
}(x) \Psi _{\rho } \;  - \; \bar{\Psi }_{\alpha } \gamma ^{\alpha
}(x)  \gamma ^{\beta }(x) \stackrel{\leftarrow}{D} _{\beta }
\gamma ^{\rho } (x) \Psi _{\rho } \; ]
$$

$$
 + \kappa \; \bar{\Psi}_{\alpha } \Psi ^{\alpha }  - {1 \over 3}
\kappa \; \bar{\Psi}_{\alpha }\;  \gamma ^{\alpha } (x)\; \gamma
_{\beta }(x) \Psi ^{\beta }  \; . \eqno(1.1)
$$

\noindent Here  $\Psi _{\alpha }$ stands for a wave function for a particle with transformation properties of local
bispinor and general covariant vector;symbols  $\rightarrow$ and $\leftarrow$
designate operators $D_{\alpha }$  acting on the right and on the left respectively
$$
\stackrel{\rightarrow}{D}_{\alpha }   =
\stackrel{\rightarrow}{\nabla} _{\alpha }  +  \Gamma _{\alpha }(x)
- i e  \; A_{\alpha }(x) \;    , \qquad
\stackrel{\leftarrow}{D}_{\alpha }   =
\stackrel{\leftarrow}{\nabla} _{\alpha }  -  \Gamma _{\alpha }(x)
+ i e  \; A_{\alpha }(x) \;    ,
$$

\noindent  $A_{\alpha }(x)$ designates a 4-potential of external electromagnetic field;  for shortness
the combination  $e/ \hbar c$ is noted as  $e$.

From Lagrangian (1.1) it follow equations for  $\Psi (x)$
and  $\bar{\Psi}(x)$:

$$
\left [ \;  [ \; \gamma^{\alpha}(x)
\stackrel{\rightarrow}{D}_{\alpha } \; + \;  \kappa \; ] \; \delta
^{\beta }_{\sigma } \; - \; {1 \over 3} \;  [  \; \gamma ^{\beta
}(x) \; \stackrel{\rightarrow}{D}_{\sigma } \; + \; \gamma
_{\sigma }(x)
 \stackrel{\rightarrow}{D}^{\beta }  \;]  \right.
$$
$$
\left. + {1 \over 3} \; \gamma _{\sigma }(x)\; [\;  \gamma ^{\alpha
}(x) \stackrel{\rightarrow}{D}_{\alpha } \; -\; \kappa \; ]  \;
\gamma ^{\beta }(x) \;  \right ] \; \Psi _{\beta }(x) = 0 \; ,
$$
$$
\eqno(1.2a)
$$

\noindent and
$$
\bar{\Psi}_{\beta}(x)\; \left [ \;  [ \;  \gamma^{\alpha} (x)
\stackrel{\leftarrow}{D} _{\alpha} \; -\; \kappa  \; ]  \; \delta
^{\beta }_{\sigma} \;  - \; {1 \over 3} \; [ \; \gamma ^{\beta
}(x) \stackrel{\leftarrow}{D}_{\sigma} \; +\;\gamma_{\sigma}(x)
\stackrel{\leftarrow}{D}^{\beta } \; ] \; \; + \right .
$$
$$
\left.  {1 \over 3} \;  \gamma ^{\beta }(x)\; [\;  \gamma ^{\alpha
}(x) \; \stackrel{\leftarrow}{D}_{\alpha}  \; + \; \kappa \; ]  \;
\gamma_{\sigma} (x) \;  \right ]  = 0   \; .
$$
$$
\eqno(1.2b)
$$

Below we use spinor representation for Dirac matrices, so we use identities

$$
\bar{\Psi}_{\beta } = \Psi ^{+}_{\beta } \; \gamma ^{0} \; ,
\qquad (\gamma ^{\beta}(x))^{+}  = \gamma ^{0} \; \gamma ^{\beta
}(x) \; , \qquad
 ( \Gamma _{\beta }(x))^{+}\; \gamma ^{0}  = -
\gamma ^{0} \; \Gamma _{\beta }(x) \; . \eqno(1.3)
$$

\noindent The order in writing operators $\gamma ^{\alpha }(x)$  and
$\stackrel{\rightarrow}{D}_{\beta }$ (also  $\gamma ^{\alpha }(x)$
and  $\stackrel{\leftarrow}{D}_{\beta }$ ) does not matter, this quantities commute with each other;
besides there exist identities

$$
 \gamma ^{\rho }(x) \; \Gamma _{\sigma }(x) \;  - \;
\gamma _{\sigma }(x) \; \Gamma ^{\rho }(x)  = \nabla _{\sigma } \;
\gamma ^{\rho }(x) \; , \eqno(1.4a)
$$
$$
\gamma ^{\rho }(x) \; \stackrel{\rightarrow}{D}_{\sigma } \; = \;
\stackrel{\rightarrow}{D}_{\sigma } \; \gamma ^{\rho }(x) \;  ,
\qquad \gamma ^{\rho }(x) \;\stackrel{\leftarrow}{D}_{\sigma }  \;
=  \; \stackrel{\leftarrow}{D}_{\sigma } \; \gamma ^{\rho }(x) \;
. \eqno(1.4b)
$$

\noindent Below we will use the formulas

$$
 \gamma ^{\alpha }(x) \; \gamma ^{\beta }(x) \; + \;
\gamma ^{\beta }(x) \; \gamma ^{\alpha }(x) =\; 2 \; g^{\alpha
\beta }(x)\; , \qquad
 \gamma ^{\alpha }\; \gamma _{\alpha } = 4  \; ,
$$
$$
\gamma ^{\alpha }(x) \; \gamma ^{\beta }(x)  =  \;
 g^{\alpha \beta }(x)
 \; + \; 2 \; \sigma ^{\alpha \beta }(x) \;   \; , \qquad
\;  \sigma ^{\alpha \beta }(x) \; = \; \sigma ^{ab} \;  e ^{\alpha
}_{(a)}(x)\;
 e^{\beta }_{(b)}(x) \; ,
$$
$$
\gamma ^{\alpha }(x) \; \gamma ^{\beta }(x) \; \gamma ^{\rho }(x)
= \;  \gamma ^{\alpha }(x) \; g^{\beta \rho }(x) \;  -  \; \gamma
^{\beta }(x) \; g^{\alpha \rho }(x) \; +
$$
$$
 \gamma ^{\rho }(x) \; g^{\alpha \beta }(x) \; + \; i \gamma ^{5}
\; \epsilon ^{\alpha \beta \rho \sigma }(x)\; \gamma _{\sigma }(x)
\;  \;  ; \eqno(1.5)
$$

\noindent they follow from the properties of usual Dirac matrices multiplied by
relevant tetrads.

Starting  with eqs.  $(1.2a,b)$, one can derive additional constraints for components of the wave function  $\Psi
_{\alpha }(x)$; thereby, in accordance with Pauli -- Fierz approach
\cite{1939-Pauli(2), 1939-Fierz-Pauli},  these constraints are  deduced from the initial lagrangian (1.1)

Indeed, let us multiply eq.  $(1.2a)$ from the left by the matrix $\gamma ^{\sigma }(x)$:

$$
\left [\;  \gamma ^{\beta } \; \gamma ^{\alpha } \;
 D_{\alpha } \; + \; \kappa \;  \gamma ^{\beta } \;  - \;
{1 \over 3} \; \gamma ^{\sigma }  \; \gamma ^{\beta } \; D_{\sigma
} \;  - \; {4 \over 3} \;  D^{\beta } \;  + \; {4 \over 3}\;
\gamma ^{\alpha } \; \gamma ^{\beta } \; D_{\alpha } \;   - \; {4
\over 3} \;  \kappa \;  \gamma ^{\beta } \; \right ] \;  \Psi
_{\beta }  = 0 \;  ,
$$

\noindent from whence it follows

$$
D_{\beta } \; \Psi ^{\beta }  = {\kappa \over 2} \;
 \gamma _{\beta } \; \Psi ^{\beta }  \; .
\eqno(1.6)
$$

\noindent It is  a first additional constraint.  Now, let us act on eq.  $(1.2a)$ from the left by operator
 $D^{\sigma }$:

$$
\left [ \;  D^{\beta } \; \gamma ^{\alpha }\; D_{\alpha }   +
\kappa  \; D^{\beta }   - {1 \over 3}  \; \gamma ^{\beta } \;
 D^{\sigma }  D_{\sigma }   -  {1 \over 3} \;  \gamma ^{\sigma } \;
D_{\sigma } \;  D^{\beta }   + \right.
$$
$$
\left.  \; {1 \over 3} \;  \gamma _{\sigma }  \; \gamma ^{\alpha }
\; D^{\sigma } D_{\alpha } \; \gamma ^{\beta }  - {\kappa \over
3}\; \gamma ^{\sigma }  D_{\sigma } \;  \gamma ^{\beta } \; \right
] \;
 \Psi _{\beta } (x)  = 0 \; .
$$

\noindent Then with the use of identity

$$
D^{\beta } \; D_{\alpha }  = \;  D_{\alpha } \; D^{\beta } \; + \;
D ^{\beta }_{\;\;\alpha } \;   , \qquad \mbox{where} \qquad D
^{\beta }_{\;\;\alpha }  =  D^{\beta } \; D_{\alpha } \; - \;
D_{\alpha } \; D^{\beta } \;,
$$

\noindent we get

$$
  \gamma ^{\alpha } \; D_{\alpha }  ( \; {2 \over  3}\;
 D^{\beta } \; - \; {\kappa \over 3}\;  \gamma ^{\beta } )\;
\Psi _{\beta } \; + \; \gamma ^{\alpha } \; D ^{\beta
}_{\;\;\alpha } \; \Psi _{\beta } \; + \; \kappa \;  D^{\beta } \;
\Psi _{\beta } \;   +  \; {1 \over 3} \;
 \sigma ^{\alpha \beta }  D_{\alpha \beta } \;
\gamma ^{\rho } \; \Psi _{\rho }   = 0 \;  .
$$

\noindent Here, the first term vanishes due to (1.6). Thus, we arrive at

$$
 -  D_{\alpha \beta }  \; \gamma ^{\alpha } \Psi ^{\beta }   +
{\kappa ^{2} \over 2} \; \gamma ^{\rho } \; \Psi _{\rho } + \; {1
\over 3} \;  \sigma ^{\alpha \beta }\;  D_{\alpha \beta }\;
 \gamma ^{\rho } \; \Psi _{\rho }   = 0  \; .
\eqno(1.7)
$$

This second additional constraint can be transformed  to the form of algebraic
relationships.  Indeed, let us detail operator  $D_{\alpha \beta}$:

$$
D_{\alpha \beta }  =  (
 \nabla _{\alpha } \; \nabla _{\beta } \; - \;
 \nabla _{\beta }  \; \nabla _{\alpha }  ) \; +
\; \hat{D}_{\alpha \beta } \;  - \; i  e \;  F_{\alpha \beta } \;
, \eqno(1.8a)
$$

\noindent  where  $F_{\alpha \beta }$ is a electromagnetic tensor;
  $\hat{D}_{\alpha \beta }$ is determined by relation

$$
\hat{D}_{\alpha \beta } =   \; \nabla _{\beta}  \Gamma _{\alpha }
- \nabla _{\beta}   \Gamma _{\beta }   +
  \Gamma _{\alpha } \; \Gamma _{\beta } \; - \; \Gamma _{\beta }\;
 \Gamma _{\alpha } \;  \;  .
\eqno(1.8b)
$$

\noindent
 With the use of definition for the bispinor connection  $\Gamma _{\alpha }$,one can produce

$$
\nabla _{\beta}  \Gamma _{\alpha }   - \nabla _{\beta}   \Gamma
_{\beta }    =
 {1 \over 2} \;  \sigma ^{ab} \; e^{\nu }_{(a)}
(   e_{(b) \nu ;\alpha ;\beta } \;  - \; e_{(b)\nu ;\beta ;\alpha
}  )
$$
$$
+  \; {1 \over 2} \;  \sigma ^{ab} \; ( e_{(a)\nu ;\alpha }\;
 e^{\nu }_{(b);\beta } \;  -\; e_{(a)\nu ;\beta }
\; e^{\nu }_{(b);\alpha }  )   \; . \eqno(1.8c)
$$

\noindent For the term  $(  \Gamma _{\alpha } \; \Gamma _{\beta } -
\Gamma _{\beta } \; \Gamma _{\alpha } )$,  using the commutative relation

$$
 [ \sigma ^{ab}, \; \sigma ^{mn}  ]  =
 (\; g^{ma}\;  \sigma ^{nb} \;  - \; g^{mb} \; \sigma ^{na}  ) \;-\;
 ( \; g^{na} \; \sigma ^{mb} \; - \; g^{nb} \; \sigma ^{ma}  ) \; ,
$$

\noindent we derive the following expression

$$
\Gamma _{\alpha } \; \Gamma _{\beta } \;  - \; \Gamma _{\beta } \;
\Gamma _{\alpha } \;   =
 - {1 \over 2}\;  \sigma ^{ab} \;
( \; e_{(a)\nu ;\alpha } \; e^{\nu }_{(b);\beta } \; -\;
       e_{(a)\nu ;\beta }  \; e^{\nu }_{(b);\alpha} \;  )\;  .
\eqno(1.8d)
$$

 \noindent Summing  $(1.8c)$ and  $(1.8d)$, we get

$$
\hat{D}_{\alpha \beta }  = {1 \over 2}\;  \sigma ^{ab} \; e ^{\nu
}_{(a)} \; (\; e_{(b)\nu ;\beta ;\alpha } \; - \;
          e_{(b)\nu ;\alpha ;\beta }\; )
$$
$$
= {1 \over 2}\; \sigma ^{ab} \; e ^{\nu }_{(a)} \; e ^{\mu
}_{(b)}\; R_{\mu \nu \beta \alpha } (x)  = {1 \over 2}\;  \sigma
^{\nu \mu }(x) \; R_{\mu \nu \beta \alpha }(x)  \; , \eqno(1.8e)
$$

\noindent where $R_{\mu \nu \beta \alpha }(x)$ stands for the Riemann tensor.
Substituting  $(1.8e)$  into   $(1.8a)$, we obtain

$$
D_{\alpha \beta }  =  \;
 (\nabla _{\alpha } \; \nabla _{\beta } \;  - \;
 \nabla _{\beta }  \; \nabla _{\alpha } ) \;  + \;
{1 \over 2}\;  \sigma ^{\mu \nu } \; R_{\mu \nu \alpha \beta } \;
- \; i e \;  F_{\alpha \beta } \;   \; . \eqno(1.9)
$$

Taking into account  (1.9), now consider (1.7). For the first term
in (1.7) we will obtain

$$
- \gamma ^{\alpha } \; D_{\alpha \beta } \; \Psi ^{\beta }      =
- \gamma ^{\alpha } \; \left [ \; (\nabla _{\alpha } \; \nabla
_{\beta } \; - \; \nabla _{\beta }  \; \nabla _{\alpha }  ) \; +
\; {1 \over 2} \;  \sigma ^{\mu \nu } \; R_{\mu \nu \alpha \beta }
\; - \; i e \;  F_{\alpha \beta }\;   \right ] \; \Psi ^{\beta }
\;  ; \eqno(1.10a)
$$

\noindent note identity

$$
- \gamma ^{\alpha } \; (\nabla _{\alpha } \; \nabla _{\beta }
-\nabla _{\beta }  \; \nabla _{\alpha }  ) \;\Psi^{\beta} = \gamma
^{\alpha }\;\Psi ^{\nu } \; R_{\nu \alpha } \; ;
$$

\noindent
 for the second term, using  (1.5), one derives

$$
-{1 \over 2} \; \gamma ^{\alpha } \; \sigma ^{\mu \nu } \;
          R_{\mu \nu \alpha \beta } \; \Psi ^{\beta }  =
 -{1 \over 4}\; \left  [ \;
\gamma ^{\alpha }  g^{\mu \nu } \; - \; \gamma ^{\mu }  g^{\alpha
\nu }\; +\; \gamma ^{\nu }  g^{\alpha \mu } \;+  \; i\; \gamma
^{5}  \epsilon ^{\alpha \mu \nu \sigma }(x)  \gamma _{\sigma } \;
\right ] \;  R_{\mu \nu \alpha \beta } \;  \Psi ^{\beta } \;  ,
$$

\noindent from whence, alowing for symmetry of the Riemann tensor
we get
 ($R_{\alpha \beta }$ is  the Ricci tensor):

$$
 - {1 \over 2} \;\gamma ^{\alpha } \; \sigma ^{\mu \nu } \;
R_{\mu \nu \alpha \beta } \; \Psi ^{\beta }    = -{1 \over 2}\;
\gamma ^{\nu } \; R_{\nu \beta } \; \Psi ^{\beta } \; .
\eqno(1.10b)
$$

\noindent This, relation  $(1.10a)$ reads

$$
- \; \gamma ^{\alpha } \;  \tilde{D}_{\alpha \beta } \; \Psi
^{\beta }  = ( \;{ 1 \over 2} \; R_{\alpha \beta } \;  +  \; ie \;
F_{\alpha \beta } ) \;
 \gamma ^{\alpha } \; \Psi ^{\beta } \; .
\eqno(1.10c)
$$

 Now, for the third term in  (1.7) we derive

$$
{1 \over 3}  \; ( \sigma ^{\alpha \beta } D_{\alpha \beta }  ) \;
\gamma^{\rho }  \; \Psi _{\rho } = {1 \over 3} \sigma ^{\alpha
\beta } \; \left [ \; ( \nabla _{\alpha }  \nabla _{\beta }   -
  \nabla _{\beta }  \nabla _{\alpha }  )  +
{1 \over 2}  \sigma ^{\mu \nu }  R_{\mu \nu \alpha \beta }    - i
e   \; F_{\alpha \beta } \;  \right ]  \gamma ^{\rho } \; \Psi
_{\rho } \; .
$$

\noindent Here the first term vanish identically (let it be  $\gamma ^{\sigma } \; \Psi _{\sigma }  = \Phi (x)$:

$$
( \; \nabla _{\alpha } \; \nabla _{\beta } \; -
           \nabla _{\beta }  \;  \nabla _{\alpha }  \;  ) \;
\Phi(x)
$$
$$
 = \partial /\partial x^{\alpha } \;
          (\;  \partial \Phi /\partial x^{\beta }\; ) \; - \;
\Gamma ^{\sigma }_{\alpha \beta } \; ( \;\partial \Phi /\partial
x^{\sigma }\; ) \; - \;
    \partial / \partial x^{\beta } \; (\; \partial \Phi /\partial x^{\alpha }\;)
\; + \;  \Gamma ^{\sigma } _{\beta \alpha } \; (\;\partial \Phi
/\partial x^{\sigma }\;) \equiv 0 \; .
$$

\noindent The second term  ($R$ is  the Ricci scalar) reads

$$
{1 \over 6} \;  \sigma ^{\alpha \beta } \; \sigma ^{\mu \nu } \;
R_{\mu \nu \alpha \beta } \; (\gamma ^{\rho } \; \Psi _{\rho })  =
-{1 \over 24} \; \gamma ^{\alpha }\; (\; \gamma ^{\beta } \;
\gamma ^{\mu } \; \gamma ^{\nu } \;) \; R_{\mu \nu \alpha \beta }
\; (\gamma ^{\rho } \; \Psi _{\rho }\; )
$$
$$
 = - {1 \over 12}\; \gamma ^{\alpha } \; \gamma ^{\beta } \;
R_{\alpha \beta } \; (\; \gamma ^{\rho }\;  \Psi _{\rho }\; ) = -
{1 \over 12} \;(g^{\alpha \beta } \; + \; 2 \; \sigma ^{\alpha
\beta }\; )\;
 R_{\alpha \beta } \; ( \; \gamma ^{\rho }\; \Psi _{\rho }\; ) =
- {1 \over 12} \;  R\; (\gamma ^{\rho }\; \Psi _{\rho }) \; .
$$

\noindent Therefore, the third term in  (1.7)  reduces to
$$
{1 \over 3}  \; ( \sigma ^{\alpha \beta } D_{\alpha \beta }  ) \;
\gamma^{\rho }  \; \Psi _{\rho } = - {1 \over 12} \;  R\; (\gamma
^{\rho }\; \Psi _{\rho }) + ie {1\over 3} \;\sigma^{\alpha \beta}
F_{\alpha \beta} \; \gamma^{\rho} \Psi_{\rho}\; .
$$

Thus, the second additional constraint (1.7) is equivalent to the algebraic relationship

$$
(\;  {1 \over 2} \; R_{\alpha \beta}  \; + \; i e \;
 F_{\alpha \beta}  \; )  \; \gamma ^{\alpha }  \;
 \Psi ^{\beta }  \; +
   \left [\;{1 \over 2} \; \kappa ^{2} \; - \;
 {1 \over 3}\;  (
\;  {1 \over 4} \; R  \; + \; i e \; F_{\alpha \beta }  \; \sigma
^{\alpha \beta } ) \;  \right  ] \;
 \gamma ^{\rho } \; \Psi _{\rho }     = \; 0 \; ;
\eqno(1.11a)
$$

\noindent for convenience let us written down the first condition as well

$$
D_{\beta } \; \Psi ^{\beta }  = {\kappa \over 2} \;
 \gamma _{\beta }  \; \Psi ^{\beta }      \; .
\eqno(1.11b)
$$

Sometime, these two relations permit us to greatly simplify
the initial wave equation   $(1.2a)$. For instance, for a free particle in Minkowski
space-time, in Cartesian coordinates and tetrad, eqs.  $(1.11 ,b)$  give

$$
\gamma ^{a} \;\; \Psi _{a}(x) = 0 \;\; , \qquad
\partial ^{a}  \; \Psi _{a}(x) = 0 \; ,
\eqno(1.12a)
$$

\noindent so that eq.  $(1.2a)$ assumes the form of four separate Dirac equations

$$
(\; \gamma ^{a} \; \partial_{a} \; + \; \kappa \;  ) \;  \Psi
_{c}(x) = 0    \; . \eqno(1.12b)
$$

Analogous situation arises in any curved space-time
with vanishing Ricci tensor. Indeed, let
$$
R_{\alpha \beta }(x) = 0 \; , \qquad  F_{\alpha \beta }(x) = 0 \;
, \eqno(1.13a)
$$

\noindent then the full systems of equations determining the particle with spin 3/2 is

$$
\gamma ^{\beta }(x) \; \Psi _{\beta }(x) = 0 \; , \;\; (\nabla
_{\beta } \;+\; \Gamma _{\beta }(x) ) \; \Psi ^{\beta }(x) = 0 \;
,
$$
$$
\left [\; \gamma ^{\alpha }(x)\; ( \nabla _{\alpha } \;+\; \Gamma
_{\alpha }(x)  )\; + \;
 \kappa \; \right   ] \; \Psi ^{\beta }(x) = 0 \; .
\eqno(1.13b)
$$

\noindent It should be noted that because $\Psi ^{\beta }(x)$ stands for a general covariant vector,
and  $\nabla _{\alpha }$   stands for a covariant derivative,  the kasr equation in $(1.13b)$
is not equivalent to four independent Dirac-like equations.

We can extend the system (1.13b) to the class of space-time model
with more general structure of the Ricci tensor
$$
R_{\alpha \beta }(x) =  {1 \over 4} \; R(x)  \; g_{\alpha \beta
}(x) \; . \eqno(1.14a)
$$

\noindent In this case, additional constraints reduce to

$$
D^{\beta }(x) \; \Psi _{\beta }(x)  =  {1 \over 2}\;
 \kappa \; \gamma ^{\beta }(x) \; \Psi _{\beta }(x) \; ,
 $$
 $$
\left (\; {1 \over 12} \; R(x)\; - \; {m^{2} \;c^{2} \over \hbar ^{2}}\;
\right ) \; [ \; \; \gamma ^{\beta }(x) \; \Psi _{\beta }(x)\; ]   = 0 \;
. \eqno(1.14b)
$$

Simplest examples of such models are de Sitter and anti de Sitter spaces.

\section{Massless field
}

Now let us specify the massless case.
It is known that in Minkowski space-time, equation for massless field  with spin 3/2
can be transformed to a special form when it become evident existence
of trivial solutions in  the form of 4-gradient of arbitrary bispinor
$$
i\gamma ^{5} \; \epsilon ^{\;\;bcd}_{a} \; \gamma _{d} \;
\partial _{b} \; \tilde{\Psi }_{c}(x)  = 0 \;\; , \qquad
\tilde{\Psi }^{0}_{c}(x) = \partial_{c} \; \psi (x) \;  .
\eqno(2.1)
$$

\noindent This property proves gauge invariance of massless wave equation, which give possibility to remove
redundant degrees of freedom .

Let us consider analogous problem in the case of a curved space-time.
It is convenient to start with the following matrix form  of eq.   $(1.2a)$
$$
\left [ \; \alpha ^{\nu }(x) \; D_{\nu } \; + \; \kappa  \;\beta
(x) \; \right ] \; \Psi (x)  = 0 \; , \eqno(2.2a)
$$
$$
\Psi (x)  = (\Psi _{\sigma }(x)) \; , \qquad (\beta )^{\;\;\sigma
}_{\rho } = \delta ^{\;\;\sigma }_{\rho } \; - \; {1 \over 3} \;
\gamma _{\rho }(x) \; \gamma ^{\sigma }(x)  \; ,
$$
$$
(\alpha ^{\nu })^{\;\;\sigma }_{\rho } = \gamma ^{\nu }(x) \;
\delta ^{\sigma }_{\rho } \; - {1 \over 3}\; \gamma ^{\sigma
}(x)\; \delta ^{\nu }_{\rho } \;
$$
$$
- \; {1 \over 3} \; \gamma _{\rho
}(x)\; g^{\nu \sigma }(x) \; + \; {1 \over 3}\; \gamma _{\rho }(x)
\; \gamma ^{\nu }(x) \; \gamma ^{\sigma }(x)  \; . \eqno(2.2b)
$$

 Let us perform two successive transformation over eq. $(2.2a)$.
 Furs, multiply it  from the left by a matrix $C$,  ant then translate equation to a new
 representation with the help of other matri $S$:

$$
\beta \; , \; \alpha ^{\nu } \;\; \Longrightarrow  \;\; \beta ' =
C \; \beta \; , \; \alpha ^{'\nu } = C \; \alpha ^{\nu } \;\;
\Longrightarrow \;\;
$$
$$
\tilde{\beta } = S \; \beta ' \; S^{-1} \; , \; \tilde{\alpha }
^{\nu } =  \; S \; \alpha ^{'\nu } \; S^{-1}\; , \;\; \tilde{\Psi
} = S \; \Psi  \; . \eqno(2.3)
$$

\noindent The relevant matrices are taken in the form

$$
C^{\;\;\beta }_{\alpha }=  \delta ^{\beta }_{\alpha } \; + \; c\;
\gamma _{\alpha }(x) \; \gamma ^{\beta }(x) \;  \; , \qquad
S^{\;\;\beta }_{\alpha } =   \delta ^{\beta }_{\alpha } \; + \; a
\; \gamma _{\alpha }(x) \; \gamma ^{\beta }(x)   \; ,
$$

$$
(S^{-1})^{\;\;\beta }_{\alpha } =   \delta ^{\beta }_{\alpha } \;
+ \; b \; \gamma _{\alpha }(x) \; \gamma ^{\beta }(x)  \; , \qquad
 a  +  b   +  4\; a b  = 0 \; .
\eqno{2.4}
$$

\noindent The quantities  $a ,\; b ,\; c $  are unknown numerical parameters;
 relationship between  $a$ and  $b$ ensures identity  $S\; S^{-1} = I$.
In accordance with  (2.3)  and    (2.4), we find  $\beta ' , \;
 \tilde{\beta }$   and  $\alpha ^{'\nu }  , \; \tilde{\alpha }^{\nu }$:

$$
(\beta ')^{\;\;\sigma }_{\rho } =  (\;
 \delta ^{\sigma }_{\rho } \;  - \;
{ c \; + \; 1 \over 3} \; \gamma _{\rho } \;\gamma ^{\sigma } ) \;
, \qquad
$$
$$
( \tilde{\beta } )^{\;\;\sigma }_{\rho }  = \{ \delta ^{\sigma
}_{\rho }\; + \;  [ \; b\;+\; ( 4 b \;+\; 1)\; ( a\; - \;(4 a\; +
\;1)\; {c\; + \;1 \over 3} ) ] \; \gamma _{\rho }\;
 \gamma ^{\sigma } \;  \} \; , \qquad
\eqno(2.5a)
$$

$$
(\alpha ^{'\nu })^{\;\;\sigma }_{\rho }  =  [ \; \gamma ^{\nu } \;
\delta ^{\sigma }_{\rho } \; - \; {1 \over 3} \; \gamma ^{\sigma
}\; \delta ^{\nu }_{\rho } \;  + (2c\;  -{1 \over 3}) \;
 \gamma _{\rho }\; g^{\nu \sigma } \;  + {1 \over 3}\;
\gamma _{\rho } \; \gamma ^{\nu }\;  \gamma ^{\sigma } ] \; ,
\qquad  \qquad
$$
$$
(\tilde{\alpha }^{\nu }) ^{\;\;\sigma }_{\rho } = \gamma ^{\nu }
\; \delta ^{\sigma }_{\rho }\; \{ \;  1\; - \;  [\;
 { b\; +\; 1 \over  3} \; + \;  \; b \;
(\; {2c \; - \; 1 \over 3} \;  (1\; + \; 4a) \;  + \; 2a  )\;
 ] \; \} \qquad \qquad
$$
$$
 +\; \gamma ^{\sigma } \; \delta ^{\nu }_{\rho } \; \{   \; { 2 b
\;-\;1 \over 3}\;+\;  [ \; { b\; + \; 1 \over 3} \; + \; b\;  (
{2c\; - \; 1 \over 3}\;
 (1 \; + \;  4a) \; + \; 2a ) \;  ]  \}  \qquad
$$
$$
 +\; \gamma _{\rho }\; g^{\nu \sigma } \;    \{ [ (2c\; - \; 1) \;
{1 \; +\; 4a \over 3} \; + \; 2a ] \; + \; [\;  {b \; + \; 1 \over
3} \; + \; b ( \; (2c \; - \; 1) \; {1 \; +\; 4a \over 3} \;  + \;
2a  ) ]  \}
$$
$$
 + \; i\; \gamma ^{5} \; \epsilon ^{\nu \sigma \mu }_{\rho } \;
\gamma _{\mu } \; [ \; {b \;+\;1 \over 3} \;+\; b \; ( \;(2c \; -
\;1)\; {1 \;+\; 4a \over 3} \; +\;  2a   ) \; ] \; .
$$
$$
\eqno(2.5b)
$$

\noindent Let us try to chose  $(a,\; b,\; c)$ so that in expression
for  $\tilde{\alpha }^{\nu }$ all terms excluding one containing Levi-Civita tensor
vanish. To this end, we must impose
restrictions

$$
a \;+\; b\; + \;4 \;a\; b  = 0 \; , \;\; 1\; - \;  [\; {b \; +\; 1
\over 3}\; + \; b \;   (\; (2c\; - \;1) \; {1\; +\; 4a \over 3} \;
+ \; 2a  )  ] = 0 \;   ,
$$
$$
{2b \; +\; 1 \over 3} \;  + \;  [ \; {b\; + \;1 \over 3}\; +\; b
\; ( \; (2c\; - \; 1) \; {1\; +\; 4a \over 3} \; +  \; 2a  )  ] =
0 \; , \;\;
$$
$$
(1 \;+\; 4a) \; {2c \; -\; 1 \over 3} \; +\; 2a\; +\; [ { b\; +\;
1 \over 3} \; + \; b\;  (\; (2c\; - \;1) \; {1 \; +\;4a \over 3}
\; + \; 2a  )  ] = 0\;  .
$$

\noindent Solution of the system is

$$
a = -  \; {1 \over 3}\; , \qquad b = - 1 \; , \qquad c = + \; 2
\eqno(2.6a)
$$

Thus, the transformation $S$ is

$$
S^{\;\;\beta }_{\alpha } =   \delta ^{\beta }_{\alpha } \; - \;
{1\over 3} \; \gamma _{\alpha }(x) \; \gamma ^{\beta }(x)   \; ,
\qquad
 \tilde{\Psi}_{\alpha} =
S^{\;\;\beta }_{\alpha }  \; \Psi_{\beta}
$$

\noindent and correspondingly in new representation the wave equation is  determined by the matrices
$$
( \tilde{\beta } )^{\;\;\sigma }_{\rho }  =
 \delta ^{\sigma }_{\rho } \; - \; \gamma _{\rho }(x) \;\gamma ^{\sigma }(x)\;
 \; ,
 $$
 $$
 ( \tilde{\alpha }^{\nu } )^{\;\;\sigma }_{\rho }  =
+ i \;\gamma ^{5} \; \epsilon ^{\;\;\nu \sigma \mu }_{\rho }(x) \;
\gamma _{\mu }(x)\;  \; . \eqno(2.6b)
$$

\noindent Expression for $\tilde{\beta }$   in  $(2.6b)$ can be  rewritten as
differently
$$
( \tilde{\beta } )^{\;\;\sigma }_{\rho } = - 2\;
       \sigma ^{\;\;\sigma }_{\rho }(x)
       $$

\noindent
 and further, with the use of identity
$$
2 \; \sigma ^{\;\;\sigma }_{\rho }(x)  = 2\; ( { 1 \over 4}\;
\gamma _{\mu }(x) ) \; \left [ \; \gamma ^{\mu }(x)\; \sigma
^{\;\;\sigma }_{\rho }(x)\; \right ] =
 \sigma ^{\;\;\sigma }_{\rho }(x) \; + \; {i \over 4}\;
\gamma _{\mu }(x) \; \gamma ^{5} \; \epsilon ^{\mu\;\;\sigma \nu
}_{\;\;\rho } \; \gamma _{\nu }(x)\;
 $$

\noindent for the matrix $\tilde{\beta }$  we get

$$
( \tilde{\beta } )^{\;\;\sigma }_{\rho }  =  {i \over 2}\;
 \gamma ^{5} \; \epsilon ^{\;\;\nu \sigma \mu }_{\rho }(x)\;
 \gamma _{\mu }(x) \; \gamma _{\nu }(x)\; .
\eqno(2.6c)
$$

Allowing for  $(2.6b,c)$,  equation for the particle with spin  $3/2$ can be presented as follows

$$
\gamma ^{5} \; \epsilon ^{\;\;\nu \sigma \mu }_{\rho }(x) \;
\gamma _{\mu }(x) \; \left [\;  i \; D_{\nu } \;  - \; {mc \over
2\hbar} \; \gamma _{\nu }(x)\; \right ]  \; \tilde{\Psi }_{\sigma
}(x)  = 0 \; .
\eqno(2.7a)
$$

\noindent At $m = 0$    we obtain an equation  (compare it with (2.1))
for massless field

$$
i\; \gamma ^{5} \; \epsilon ^{\;\;\nu \sigma \mu }_{\rho }(x)\;
\gamma _{\mu }(x)\; \left [ \; \nabla _{\nu }\; + \; \Gamma _{\nu
}(x)  \; \right ]  \; \tilde{\Psi }_{\sigma }(x) = 0 \;  .
\eqno(2.7b)
$$

Not let us investigate the problem of possible existence of  solutions in the form of 4-gradient of arbitrary bispinor field.
Substituting the function $\tilde{\Psi }^{0}_{\sigma }(x)$  of the form

$$
\tilde{\Psi }^{0}_{\beta }(x) =  [\;
 \nabla _{\beta }\; + \; \Gamma _{\beta }(x)\; ]\;
 \Psi (x) \; ,
\eqno(2.8a)
$$

\noindent into eq. $(2.7b)$, we get

$$
{i \over 2} \; \gamma ^{5} \; \epsilon ^{\;\;\nu \sigma \mu
}_{\rho }(x)\; \gamma _{\mu }(x) \; [\; D_{\nu }\; ,\; D_{\sigma }
]_{-}\; \Psi  (x)  = 0 \; .
$$
Taking into account expression (1.9) for the commutator
$[ D_{\nu } ,\; D_{\sigma }]_{-}$  when  $F_{\nu \mu}=0$, and also allowing for that
the bispinor  $\Psi$ is a scalar in general covariant sense, we get
$$
{i \over 4} \; \gamma ^{5} \; \epsilon ^{\;\;\nu \sigma \mu
}_{\rho }(x) \; \gamma _{\mu }(x)\;  [\; \sigma ^{\alpha \beta
}(x) \; R_{\alpha \beta \nu \sigma }(x) \; ] \; \Psi  (x) = 0 \; .
$$

\noindent Further, we obtain

$$
{i \over 4}\; \gamma ^{5} \; \epsilon ^{\;\;\nu \sigma \mu }_{\rho
}(x) \; \left [\; \gamma ^{\beta }(x) \; R_{\mu \beta \nu \sigma
}(x) \;+\; {i \over 2}\; \gamma ^{5} \; \epsilon ^{\;\;\alpha
\beta s}_{\mu }(x) \; \gamma _{s}(x) \; R_{\alpha \beta \nu \sigma
}(x) \; \right  ]\; \Psi (x) = 0 \; ;
$$

\noindent therefore   arrive at

$$
 R_{\alpha \beta \nu \sigma }(x) \;
[\; \epsilon ^{\;\;\nu \sigma \mu }_{\rho }(x)\; \epsilon ^{\alpha
\beta s}_{\;\;\;\;\;\;\mu }(x)\; ]\;  [\;  \gamma _{s}(x) \; \Psi (x)
) \; ] \;\;
 = 0 \; .
\eqno{2.8b}
$$

\noindent Using the known formula
$$
\epsilon ^{\;\;\nu \sigma \mu }_{\rho }(x) \; \epsilon ^{\alpha
\beta s}_{\;\;\;\;\;\;\mu }(x) =  \mbox{det} \; \left |
\begin{array}{ccc}
\delta ^{\alpha }_{\rho } & \delta ^{\beta }_{\rho } & \delta ^{s}_{\rho } \\
g^{\nu \alpha }(x)    &    g^{\nu \beta }(x)    &   g^{\nu s}(x)   \\
g^{\sigma \alpha }(x) &   g^{\sigma \beta }(x)  &   g^{\sigma
s}(x) \noindent \end{array} \right |   \; ,
$$

\noindent from $(2.8b)$ we derive relation needed

$$
 [\; R_{\alpha \beta }(x)\; - \; {1 \over 2}\; R(x)\;
 g_{\alpha \beta }(x) ] \;\;   \gamma ^{\beta }(x) \; \Psi (x) = 0 \; .
\eqno(2.8c)
$$

Thus, we conclude that in the region where
$$
R_{\alpha \beta }(x)\; - {1 \over 2}\; R(x)\;
g_{\alpha \beta }(x)  = 0 \; ,
$$

 \noindent the massless particle with spin 3/2 possess a gauge symmetry and thereby in such regions
 it is a correctly defined massless object;
 otherwise it is not clear how one can determine a massless field.

\end{document}